\documentclass[11pt,letterpaper,notoc]{JHEP3}
\usepackage{amsmath,cite,feynmp}

\newcommand{\nn}{\nonumber}
\def\ltap{\ \raise.3ex\hbox{$<$\kern-.75em\lower1ex\hbox{$\sim$}}\ }
\def\gtap{\ \raise.3ex\hbox{$>$\kern-.75em\lower1ex\hbox{$\sim$}}\ }

\newcommand{\tr}{{\rm tr}\,}
\newcommand{\grad}{\nabla}

\newcommand{\OO}{\mathcal{O}}

\newcommand{\MM}{\mathcal{M}}
\newcommand{\LL}{\mathcal{L}}
\newcommand{\NN}{\mathcal{N}}

\newcommand{\sslash}[1]{\ensuremath \raisebox{0.025cm}{\slash}\hspace{-0.21cm}#1\/}
\newcommand{\be}{\begin{eqnarray}}
\newcommand{\ee}{\end{eqnarray}}
\newcommand{\bea}[1]{\left(\begin{array}{#1}}
\newcommand{\ena}{\end{array}\right)}

\begin{fmffile}{graphs}

\title{On-Shell Recursion Relations for Generic Theories}

\author{Clifford Cheung$^{1,2}$\\
$^1$Jefferson Laboratory of Physics, Harvard
  University, Cambridge, MA 02138 \\
$^2$School of Natural Sciences,
  Institute for Advanced Study, Princeton, NJ 08540 \\
 email:
  \email{cwcheung@fas.harvard.edu}}

\abstract{We show that on-shell recursion relations hold for tree
amplitudes in generic two derivative theories of multiple particle
species and diverse spins. For example, in a gauge theory coupled
to scalars and fermions, any amplitude with at least one gluon
obeys a recursion relation. In (super)gravity coupled to scalars
and fermions, the same holds for any amplitude with at least one
graviton. This result pertains to a broad class of theories,
including QCD, $\NN =4$ SYM, and $\NN=8$ supergravity.}

\begin{document}

\unitlength=1mm

\section{Introduction}

There is gathering evidence that on-shell amplitudes are far
simpler than one would naively expect from conventional quantum
field theory. This observation goes back more than two decades,
when Parke and Taylor \cite{Parke:1986gb} showed that tree-level
maximally helicity violating (MHV) gluon amplitudes take an
incredibly simple form. In recent years further progress has been
made, particularly regarding amplitudes in gauge and gravity
theories. Largely inspired by Witten's twistor formulation of 4D
Yang-Mills \cite{Witten:2003nn}, techniques like the CSW rules
\cite{Cachazo:2004kj} and the BCFW recursion relations
\cite{Britto:2004ap,Britto:2005fq} have provided a better
theoretical understanding of gauge theories, as well as a
practical toolbox of methods for calculating amplitudes. Since
then, recursion relations have also been derived for gauge
theories with massive particles \cite{Badger:2005zh} and pure
gravity
\cite{Cachazo:2005ca,Bedford:2005yy,BjerrumBohr:2005jr,Benincasa:2007qj}.


In a nutshell, the BCFW recursion relations are a way of writing
on-shell tree amplitudes as a sum over products of lower point
on-shell tree amplitudes evaluated at complex momenta.
 Calculationally, they are an incredibly efficient method for
 computing amplitudes since they directly relate physical
amplitudes, making no reference to an underlying Lagrangian or the
machinery of Feynman diagrams. Indeed, despite their usual
versatility, Feynman diagrams simply become too cumbersome and
numerous to be effective when the number of external legs becomes
large. Moreover, Feynman diagrams are, arguably, a somewhat
un-physical representation of multi-particle scattering since they
are not even individually gauge invariant. In contrast, the
recursion relations make no reference to off-shell data, and are
indicative of a completely on-shell (albeit complexified) S-matrix
formulation of quantum field theory.

Ultimately, recursion relations are possible because tree
amplitudes are rational functions of the external momenta.  Thus,
given a complex deformation of the momenta parameterized by a
complex number $z$, the amplitude will be a meromorphic function
of $z$. Assuming furthermore that the amplitude vanishes as
$z\rightarrow \infty$, then it is characterized entirely by its
poles. Since each pole can be thought of as a factorization
channel of the amplitude, the residue at each pole is simply a
product of lower point amplitudes.  The bottom line is that
vanishing large $z$ behavior is a sufficient condition for the
existence of recursion relations.

 In this paper we argue that this criterion is satisfied by tree
 amplitudes in two derivative gauge
and (super)gravity theories coupled to scalars and fermions. For
example, in a theory of spin $\leq 1$, any amplitude with at least
one gluon can be recursed, while in a theory of spin $\leq 2$,
this is true of any amplitude with at least one graviton. Some
notable examples for which this holds are QCD, $\NN=4$ SYM and
$\NN=8$ supergravity. Our proof follows the approach of
\cite{ArkaniHamed:2008yf}.

The outline of the paper is as follows.  In section 2 we review
the derivation of recursion relations for a generic tree
amplitude. We show that vanishing large $z$ behavior implies that
an amplitude can be recursed.  In sections 3 and 4 we argue that
this criterion is satisfied in a broad class of amplitudes in
gauge and (super)gravity theories coupled to scalars and fermions.
We conclude in section 5.

\section{A Proof of Recursion Relations}

The BCFW recursion relations were originally proven for Yang-Mills
theory in $D=4$ dimensions using the spinor helicity formalism
\cite{Britto:2004ap,Britto:2005fq}. The proof was later extended
to arbitrary $D$ in \cite{ArkaniHamed:2008yf}, where it was also
emphasized that recursion relations are a generic property of
tree-level amplitudes that vanish at large complex momenta. In
this section we review these arguments, keeping the discussion
general and assuming nothing about the particle content or
interactions of the underlying theory.

 To begin, consider a tree
amplitude with $N+2$ external legs in $D$ dimensions. We label two
of the legs by 1 and 2, and their momenta by $p_1$ and $p_2$,
respectively. We label the other $N$ momenta by $k_i$. It is
possible to deform $p_1$ and $p_2$ in a complex momentum direction
$q$ while still maintaining momentum conservation:
\be p_1(z) = p_1 + z q  \\ p_2(z) = p_2 - z q  \nn \ee
where $z$ is a complex parameter.  We define $q^2 =q  \cdot
p_{1,2} = 0$, so that $p_{1,2}(z)$ are still on-shell momenta,
albeit complex momenta. Since our main concern is the large $z$
behavior of amplitudes, it is natural to refer to $p_{1,2}(z)$ as
``hard'' and
 $k_i$ as ``soft''.

Applying the Cauchy's residue theorem, we know that
\be \oint \frac{\MM(z)}{z} &=& \sum_{z_I} \textrm{Res}
\left(\frac{\MM(z_I)}{z_I} \right) = 0 \ee where $z_I$ label the
poles of $\MM(z)/z$.  If $\MM(z) \rightarrow 0$ as $z\rightarrow
\infty$, then there is no pole an infinity and the sum is only
over residues at finite $z_I$.  From the pole at $z=0$ we obtain
$\MM(0)$, which we immediately recognize as the tree amplitude at
real external momenta. Since tree amplitudes are rational
functions of the external momenta, we know that the rest of the
poles occur when a sum of external momenta go on-shell.  To see
this explicitly, let us partition all the soft momenta into two
groups $I_1$ and $I_2$, associated with particles 1 and 2,
respectively. We assume these groups are of size $N_1$ and $N_2$,
so $N_1 + N_2 =N$.  We label the soft momenta by $k_{i_1}$ and
$k_{i_2}$, where $i_1 \in I_1$  and $i_2 \in I_2$, and the sum of
the soft momenta in each group by $K_{I_1} = \sum k_{i_1}$ and
$K_{I_2} = \sum k_{i_2}$. In this language, the condition of
momentum conservation becomes $p_1(z)+p_2(z) + K_{I_1} + K_{I_2}
=0$. Next, consider the pole that occurs when $p_1(z) + K_{I_1}$
goes on-shell. We denote the value of $z$ at this pole by
$z_{I_1}$.  Near $z_{I_1}$, the amplitude factorizes into two
lower point amplitudes evaluated at $z_{I_1}$:
\be \MM^{N+2}(p_{1,2}(z),k_{i}) &\rightarrow& \sum_h
\MM^{N_1+2}(p_1(z),k_{i_1},h)\frac{1}{(p_1(z) + K_{I_1})^2}
\MM^{N_2+2}(p_2(z),k_{i_2},-h) \ee where we have included
superscripts that label the number of external legs in each
amplitude. Here $h$ sums over the species and polarization of the
intermediate particle going on-shell. Thus, the sum over poles is
equivalently a sum over factorization channels of the amplitude,
where the residue at each pole is the product of lower point
amplitudes evaluated at the pole. Rewriting the sum over poles as
a sum over the partitions $I_1$ and $I_2$, we obtain the recursion
relation
\be \MM^{N+2}(p_{1,2},k_{i}) &=& \sum_{I_1,h}
\MM^{N_1+2}(p_1(z_{I_1}),k_{i_1},h)\frac{1}{(p_1 + K_{I_1})^2}
\MM^{N_2+2}(p_2(z_{I_1}),k_{i_2},-h) \ee
In diagrammatic form, it is
 \be
\parbox{40mm} {
\begin{fmfgraph*}(40,30)
\fmfbottom{p1,p2} \fmftop{k1,k2,k3,k4,k5,k6} \fmfblob{.2w}{q}
\fmf{}{p1,q} \fmf{}{p2,q} \fmf{}{k1,q} \fmf{}{k6,q} \fmffreeze
\fmf{}{k2,q} \fmf{}{k3,q} \fmf{}{k4,q} \fmf{}{k5,q}
\fmflabel{$p_1$}{p1} \fmflabel{$p_2$}{p2}
\end{fmfgraph*}} \quad\quad &=& \quad
  \sum_{I_1 \cup I_2, h}
  \quad
\parbox{40mm}{ \begin{fmfgraph*}(40,30)
\fmfbottom{p1,p2} \fmftop{k1,k2,k3,k4,k5,k6} \fmfblob{.2w}{q1}
\fmfblob{.2w}{q2} \fmf{,label=$h$}{q1,q2} \fmf{}{p1,q1}
\fmf{}{p2,q2} \fmf{}{k1,q1} \fmf{}{k6,q2} \fmffreeze \fmf{}{k2,q1}
\fmf{}{k3,q1} \fmf{}{k4,q2} \fmf{}{k5,q2} \fmflabel{$p_1$}{p1}
\fmflabel{$p_2$}{p2} \fmflabel{$I_1$}{k2}\fmflabel{$I_2$}{k5}
\end{fmfgraph*}}
\ee where $h$ sums over species and polarizations. We have argued
that the above recursion relations hold as long as the pole at
infinity is absent, i$.$e$.$ as long as $\MM(z\rightarrow \infty)
=0$. Naively, one would expect this criterion to fail for gauge
and gravity amplitudes, since they contain derivative
interactions. However, in \cite{ArkaniHamed:2008yf} it was shown
that the opposite is true---an enhanced ``spin Lorentz symmetry''
yields vanishing large $z$ behavior in pure gauge theories and
even better behavior in gravity!  In this paper we show that
amplitudes in generic two derivative gauge and gravity theories
also satisfy the criterion that $\MM(z\rightarrow \infty)=0$. We
begin with theories of spin $\leq 1$.


\section{Spin $\leq 1$ Amplitudes}

In this section we consider a generic gauge theory coupled to
scalars and fermions.  We argue that $\MM(z)\rightarrow 0$ as
$z\rightarrow \infty$ for any amplitude with at least one gluon.
Consequently, any such amplitude obeys an on-shell recursion
relation. Our reasoning is independent of any particular choice of
charges for the matter fields, but requires that the theory be
limited to two derivative interactions.  For example, our result
holds for QCD and $\NN = 4$ SYM.

Our approach follows closely that of \cite{ArkaniHamed:2008yf}.
There the authors present a particularly nice physical
interpretation for the large $z$ behavior of amplitudes, which we
now review. To begin, consider $\MM(0)$, the tree amplitude at
real external momenta, in the limit where $p_{1,2}$ are very hard.
Taking particles 1 and 2 to be incoming and outgoing, we can
interpret this process as a hard particle shooting through a haze
of soft particles. In this eikonal limit, all of the soft dynamics
can parameterized by a classical background through which the hard
particle propagates. Thus, to determine the large momentum scaling
of the amplitude, it suffices to compute the two point function of
hard fluctuations in a soft background. This intuition persists in
the case where $p_{1,2}$ are complexified to $p_{1,2}(z)$.  Here
the hard limit is defined by $z\rightarrow \infty$, but like
before, all of the $z$-independent soft physics goes into
determining some classical background. Thus, the bottom line is
that the large $z$ behavior of an amplitude can be determined by
using the background field method.  Our proof occurs in three
steps:
\begin{itemize}
\label{checklist}
\item \textbf{Expand around a background: } Expand the action in terms of hard fluctuations around a soft background.  To evaluate $\MM(z\rightarrow \infty)$,
simply take the large $z$ limit of the two point function of
fluctuations in this background.
\item \textbf{Choose a gauge to remove derivative interactions: } Derivatives from gauge interactions
naively spoil large $z$ behavior. Remove large $z$ contributions
from these interactions by going to light-cone gauge for the
background ($q \cdot A$=0) and choosing an appropriate $R_\xi$
gauge for the fluctuation. Note that background light-cone gauge
cannot be chosen for certain ``unique diagrams,''
\cite{ArkaniHamed:2008yf} so these diagrams must be checked
explicitly.
\item \textbf{Check remaining diagrams:}  Choosing a gauge removes
most derivative interactions, but we still have to check a small
number of diagrams explicitly.  These include i) diagrams with no
hard propagators (of which the unique diagrams are a subset), and
ii) diagrams with only hard fermion propagators.  Because the
enhanced spin Lorentz symmetry of pure gauge theory
\cite{ArkaniHamed:2008yf} persists in a gauge theory coupled to
scalars and fermions, we find that these contributions all vanish
at large $z$.

\end{itemize}


\subsection{Background Field Lagrangian}

For concreteness, let us consider Yang-Mills theory minimally
coupled to an adjoint scalar and a fundamental fermion.  Our
arguments will not depend on this particular choice of charges.
The action is
\be \mathcal{L} &=& -\frac{1}{2} \tr F_{\mu\nu} F^{\mu\nu} + \tr
D_\mu \Phi D^\mu \Phi + \bar{\Psi} i \sslash{D} \Psi + \lambda
 \bar{\Psi}\Phi \Psi\ee Next, let us expand around a background for every
field, using lowercase/uppercase to denote
fluctuations/backgrounds:
\be A_\mu &\rightarrow& A_\mu + a_\mu  , \\
\Psi_\alpha &\rightarrow& \Psi_\alpha + \psi_\alpha  , \nn  \\
\Phi &\rightarrow& \Phi + \phi \nn \ee Expanding in powers of the
gluon fluctuations yields
\be \mathcal{L} &=& \LL_{(0)} + \LL_{(1)}+\LL_{(2)}+ \ldots \\
\LL_{(2)} &=& -\frac{1}{2} \tr D_{[\mu} a_{\nu]} D^{[\mu} a^{\nu]}
- i \tr [a_\mu,a_\nu] F^{\mu\nu}  - \tr [a_\mu,\Phi]
[a_\nu,\Phi]\eta^{\mu\nu} \nn \\
 \LL_{(1)} &=& \tr a_\mu J^\mu_{(1)} \nn \\ &=&
2\tr a_\mu\left([\Phi,iD^\mu \phi]+ [\phi, D^\mu \Phi]\right) +
\bar{\Psi}\sslash{a} \psi + \bar{\psi}\sslash{a} \Psi \nn \ee
%
%
where $D_\mu$ is a background gauge covariant derivative and
$J^\mu_{(1)}$ is the gauge current expanded to linear order in the
scalar and fermion fluctuations.

The above action only contains terms that are quadratic in the
fluctuations $a$, $\psi$, and $\phi$. This is because we have made
the very important assumption that tadpoles in the fluctuation
vanish, i$.$e$.$ the background fields obey their equations of
motion. In terms of amplitudes this corresponds to putting soft
external legs on-shell, which is of course necessary if on-shell
recursion relations are to hold.

\subsection{Eliminating $\OO(z)$ Vertices}

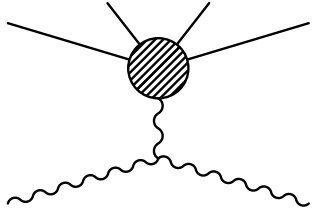
\begin{figure}[t]
\label{fig:unique} \be
\parbox{40mm} {
\begin{fmfgraph*}(40,30)
\fmfbottom{p1,p2} \fmftop{k1,k2,k3,k4} \fmfblob{.2w}{q1}
\fmf{photon}{q1,q2} \fmf{photon}{p1,q2} \fmf{photon}{p2,q2}
 \fmf{}{k1,q1} \fmf{}{k4,q1}
\fmffreeze \fmf{}{k2,q1} \fmf{}{k3,q1}
\end{fmfgraph*}} \nn
\ee \caption{The unique diagram.  Since gluons 1 and 2 meet
directly at a vertex, the momentum flowing into the soft gluon is
$p_1 + p_2$.  For this reason, background field light-cone gauge
cannot be chosen for this diagram.}
\end{figure}

In order to determine $\MM(z)$, we simply compute the two point
function of hard fluctuations in the presence of the soft
background.  We then take the large $z$ limit. Since we are
concerned with large $z$ behavior,  our first worry is
interactions with derivatives acting on the hard fields, $a$,
$\psi$ and $\phi$. Naively, these derivatives generate powers of
$z$ that blow up as $z\rightarrow \infty$. Since these terms only
show up in $D_{[\mu} a_{\nu]}D^{[\mu} a^{\nu]}$, $D_\mu\phi
D^\mu\phi$, and the mixing term, $a_\mu [\Phi,iD^\mu \phi]$, the
dangerous terms are all proportional to either $q \cdot A$ or $q
\cdot a$. This statement actually holds for any two derivative
theory, renormalizable or not.

Terms with $q$ dotted into the background gluon can be eliminated
by choosing background light-cone gauge, that is $q \cdot A
=0$.\footnote{In higher derivative theories, for example
Euler-Heisenberg theory, there are additional $\OO(z)$ terms of
the form $q_\mu F^{\mu\nu}$ that cannot be removed by background
light-cone gauge.  For this reason we restrict to two derivative
theories.}
However, as shown in \cite{ArkaniHamed:2008yf}, it is actually
impossible to choose this particular gauge for a class of
so-called unique diagrams. To see this, consider the gauge choice
$A_\mu \rightarrow A_\mu ' $, where $q \cdot A' =0$ and
\be A_\mu' &=& A_\mu + p_\mu \Omega \ee Clearly, $\Omega$ becomes
singular if $q \cdot p = 0$, which only happens if $p = p_1 +
p_2$, corresponding to the unique diagram (see figure 1)  in which
particles 1 and 2 meet directly at a trilinear vertex with a soft
gluon. Since we are only concerned with diagrams in which particle
1 is a gluon, the only unique diagram occurs when particle 2 is
also a gluon, simply because there is no interaction between two
gluons and a non-gluon\footnote{This is actually not true for all
two derivative theories. In the presence of an axion or a singlet
scalar, there can be operators of the form $a
F_{\mu\nu}\tilde{F}^{\mu\nu}$ and $b F_{\mu\nu}F^{\mu\nu}$, which
introduce new unique diagrams.  We assume that such interactions
are absent.}. This unique diagram arises in pure gauge theory, and
was shown to vanish at large $z$ \cite{ArkaniHamed:2008yf}.

Next, let us consider terms involving $q$ dotted into the
fluctuation. These terms arise from $(D_{\mu}a^\mu)^2$ and the
mixing term.  We recognize the latter as simply the mixing term
between gauge bosons and goldstone bosons in a gauge theory with
spontaneous symmetry breaking.  Consequently, both terms can be
removed in the usual way by an appropriate $R_\xi$ gauge choice
for the fluctuation:
\be \LL_\xi &=& \frac{1}{\xi} \tr (D_\mu a^\mu + i \xi
[\Phi,\phi])^2 \ee
For the choice of $\xi =1$, all derivative interactions are
removed, and the quadratic gluon action becomes
\be \label{eq:gluon} \ \LL_{(2)} &=& -\frac{1}{2} \tr \eta^{ab}
D_{\mu} a_{a} D^{\mu} a_{b} - i \tr [a_a,a_b] F^{ab}  - \tr
[a_a,\Phi] [a_b,\Phi]\eta^{ab}
 \ee
where we have re-written some Greek indices as Latin indices in
order to emphasize the spin Lorentz symmetry, which is nothing
more than the fact that the kinetic term enjoys an enhanced
Lorentz symmetry that acts only on the $ab$ indices.  In the
$z\rightarrow\infty$ limit, this term dominates over everything,
so the spin Lorentz symmetry is a symmetry of the leading order in
$z$ contribution. As discussed in \cite{ArkaniHamed:2008yf}, this
spin Lorentz symmetry is necessary to show that pure gauge theory
amplitudes vanish at large $z$.

\subsection{Checking Explicit Diagrams}

Any Feynman diagram is simply a product of interaction vertices
and propagators.  Since our choice of gauge has fixed every
interaction to go as $\OO(1)$, the only question is how
propagators scale at large $z$. A hard boson propagator goes as
$1/(p+zq)^2 = \OO(1/z)$, while a hard fermion propagator actually
scales as $\OO(1)$:
\be \frac{1}{\sslash{p}(z)} &=& \frac{\sslash{p} + z
\sslash{q}}{p^2 + 2 z p \cdot q} \overset{z\rightarrow\infty}{=}
\frac{\sslash{q}}{2 p \cdot q}\ee
Thus, any diagram with at least one hard boson propagator will
necessarily vanish at large $z$.  In contrast, i) any diagram with
no hard propagators, and ii) any diagram with only hard fermion
propagators will naively go as $\OO(1)$ at large $z$. In this
section we explicitly check that these two classes of diagrams do
not spoil the large $z$ behavior.

To begin, we assume that particles 1 and 2 are both gluons,
leaving the mixed case for a later section.  First, let us
consider all diagrams with no hard propagators (see figures
2a-2c), all of which have the structure that particles 1 and 2
meet directly at a vertex. Consequently, these diagrams only
contain one interaction vertex involving hard fields, and it can
be read directly off the quadratic gluon action in equation
\ref{eq:gluon}. Since the action differs from that of a pure gauge
theory simply by $\tr [a_a,\Phi][a_b,\Phi]\eta^{ab}$, the two
point function of hard gluons gets an additional contribution
\be \delta \MM^{ab}_{1} &=& A\eta^{ab} \ee
on top of the pure gauge theory, where here $A$ parameterizes the
soft physics of the scalar background.

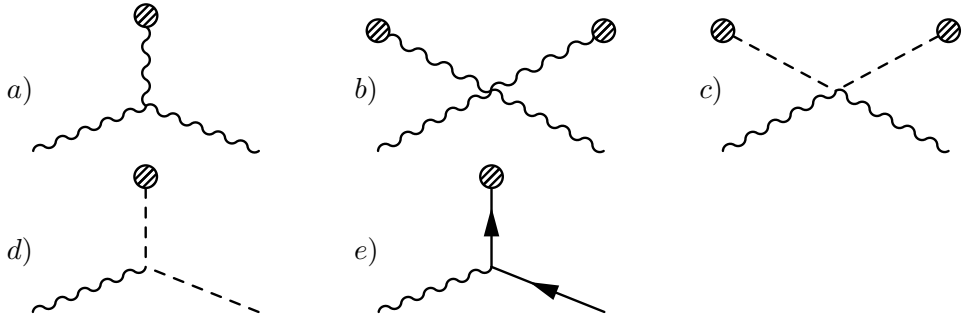
\begin{figure}[t]
\label{fig:nohardprop} \be && a) \parbox{35mm} {
\begin{fmfgraph*}(30,20)
\fmfbottom{p1,p2} \fmftop{k1} \fmf{photon}{p1,q}
 \fmf{photon}{p2,q}
\fmf{photon}{k1,q} \fmfblob{.1w}{k1}
\end{fmfgraph*}} \qquad
b) \parbox{35mm} {
\begin{fmfgraph*}(30,20)
\fmfbottom{p1,p2} \fmftop{k1,k2} \fmf{photon}{p1,q}
 \fmf{photon}{p2,q}
\fmf{photon}{k1,q} \fmf{photon}{k2,q}
\fmfblob{.1w}{k1}\fmfblob{.1w}{k2}
\end{fmfgraph*}} \qquad
c) \parbox{35mm} {
\begin{fmfgraph*}(30,20)
\fmfbottom{p1,p2} \fmftop{k1,k2} \fmf{photon}{p1,q}
 \fmf{photon}{p2,q}
\fmf{dashes}{k1,q} \fmf{dashes}{k2,q}
\fmfblob{.1w}{k1}\fmfblob{.1w}{k2}
\end{fmfgraph*}} \nn \\
&& d) \parbox{35mm} {
\begin{fmfgraph*}(30,20)
\fmfbottom{p1,p2} \fmftop{k1} \fmf{photon}{p1,q}
 \fmf{dashes}{p2,q}
\fmf{dashes}{k1,q} \fmfblob{.1w}{k1}
\end{fmfgraph*}} \qquad
e) \parbox{35mm} {
\begin{fmfgraph*}(30,20)
\fmfbottom{p1,p2} \fmftop{k1} \fmf{photon}{p1,q}
 \fmf{fermion}{p2,q}
\fmf{fermion}{q,k1} \fmfblob{.1w}{k1}
\end{fmfgraph*}} \nn
\ee \caption{All diagrams with no hard propagators.  The blobs
represent insertions of a classical background that parameterizes
all of the soft physics.  In diagrams $a)-c)$ particles 1 and 2
are both gluons, while $d)$ and $e)$ are mixed diagrams. Diagrams
$a)$ (the unique diagram) and $b)$ occur in pure gauge theory;
after dotting into the appropriate polarizations, they vanish at
large $z$ \cite{ArkaniHamed:2008yf}. Diagram $c)$ is proportional
to $\eta^{ab}$, so it preserves the spin Lorentz symmetry and is
$\OO(1/z)$ after dotting into polarizations. Diagrams $d)$ and
$e)$ are $\OO(1/z)$ after dotting into polarizations.}
\end{figure}

Next, consider the contribution from diagrams with only hard
fermion propagators. Since our gauge choice leaves only $\OO(1)$
interactions and hard fermion propagators go as $\OO(1)$, these
diagrams naively contribute at $\OO(1)$. However, this leading
order piece actually vanishes! To see this, we observe that the
leading order in $z$ contribution is obtained by taking a
$\sslash{q}$ from every hard propagator numerator. Excluding even
one $\sslash{q}$ introduces a factor of $1/z$, yielding a
subleading contribution. For this reason, any helicity flipping
insertions, such as masses or Yukawas, contribute only at
$\OO(1/z)$.

Consequently, for the leading order contribution, the only allowed
interactions along the fermion line are gauge interactions.  Thus,
the corresponding Feynman diagram is comprised of alternating
insertions of gluons and $\sslash{q}$ terms (see figure 3):
\be \MM^{a c_1 \ldots c_n b} &\sim& \gamma^{a}
\sslash{q}\gamma^{c_1} \sslash{q}\ldots \sslash{q} \gamma^{c_{n}}
\sslash{q} \gamma^b
\\ &\sim& (q^{c_1} \ldots q^{c_{n}}) \gamma^{a} \sslash{q} \gamma^{b} \nn \ee
where here $a$ and $b$ label the hard gluons at either end of the
fermion line, and in the second line we have anti-commuted gamma
matrices and used that $\sslash{q} \sslash{q} = q^2 =0$.  Without
loss of generality we can split $\MM^{a c_1 \ldots c_n b}$ into
components that are symmetric and anti-symmetric in $a$ and $b$.
The symmetric piece is proportional to $\gamma^{(a} \sslash{q}
\gamma^{b)} = 2 q^{(a} \gamma^{b)} - 2 \eta^{ab} \sslash{q}$.
Finally, after dotting the $c_i$ into soft gluon polarization
vectors and sandwiching the whole expression between soft fermion
polarization spinors, we obtain the contribution to the two point
function of hard gluons:
\be \label{eq:fermion} \delta\MM^{ab}_{2} &=& B^{[ab]} + q^{(a}
C^{b)}- \eta^{ab}(q\cdot C)\ee
where $B^{[ab]}$ and $C^a$ are functions of the soft backgrounds
and $B^{[ab]}$ is anti-symmetric.


Summing the contributions from diagrams with no hard propagators
and diagrams with only hard fermion propagators, we find that the
full amplitude becomes
\be \label{eq:gauge} \MM^{ab} &=& \MM^{ab}_{\rm gluon} + \delta\MM^{ab}_{1} + \delta\MM^{ab}_{2} \\
&=& \MM^{ab}_{\rm gluon} + A \eta^{ab} + B^{[ab]} + q^{(a} C^{b)}-
\eta^{ab}(q\cdot C) + \OO(1/z) \nn\ee
where $\MM^{ab}_{\rm gluon}$ is the contribution from the pure
gauge theory, which was calculated in \cite{ArkaniHamed:2008yf}.

Dotting into polarizations, the amplitude becomes $\MM =
\epsilon_{1a}^- \MM^{ab} \epsilon_{2b}$, where without loss of
generality we have defined gluon 1 to have negative helicity and
gluon 2 to be arbitrary.  In $D=4$ dimensions, $q$ is basically
the same as the polarization vectors for the real momenta $p_1$
and $p_2$; in particular, $\epsilon_1^- = \epsilon_2^+ = q$ and
$\epsilon_1^+ = \epsilon_2^- = q^*$.  This of course makes sense
because $\epsilon_{1,2}$ and $q$ obey the same defining equations,
$q^2 =q \cdot p_{1,2} = 0$.  If we now complex deform $p_{1,2}
\rightarrow p_{1,2}(z)$, then the polarizations must be modified
appropriately to remain normalized to unity and orthogonal to
$p_{1,2}(z)$. Given these constraints the polarizations take the
form \cite{ArkaniHamed:2008yf}
\be \epsilon_{1a}^{-} &=& q_a\overset{\rm gauge}{=}  -\frac{p_{1a}}{z} \\
\epsilon_{2a}^{\pm} &=& \left\{ \begin{array}{cc} q_a, &
\textrm{$(+)$} \\
q_{a}^*+z p_{1a}, & \textrm{$(-)$}
  \end{array}
  \right. \nn
\ee
where $\epsilon_1^- = q$ is gauge equivalent to $\epsilon_1^- =
-p_1/z$ because they are related by a gauge transformation:
\be \epsilon_{1\mu}^-&\rightarrow& \epsilon_{1\mu}^- +
p_{1\mu}(z)\left(- \frac{1}{z}\right) \ee
 where the gauge transformation of course involves the
complexified momentum.  We note that in $D>4$ dimensions, there
are an additional $D-4$ $z$-independent polarizations $\epsilon^T$
which span the vector space orthogonal to $p_{1,2}$, $q$ and
$q^*$.

By dotting polarizations into $\MM^{ab}$, we find that the
$(-,+)$, $(-,-)$, and $(-,T)$ amplitudes are
\be \MM^{-,+} &=& q_a \MM^{ab} q_b \\ &=& q_a (A \eta^{ab} +
B^{[ab]} +
q^{(a} C^{b)}- \eta^{ab}(q\cdot C))q_b + \OO(1/z) \nn \\ &\rightarrow& \OO(1/z) \nn \\
\MM^{-,-} &=& -\frac{1}{z} p_{1a} \MM^{ab} (q_b^*+z p_{1b})\\
& =& -p_{1a} (A \eta^{ab} + B^{[ab]} + q^{(a} C^{b)}-
\eta^{ab}(q\cdot C)) p_{1b} + \OO(1/z) \nn
\\&\rightarrow& \OO(1/z) \nn \\
\MM^{-,T} &=& -\frac{1}{z} p_{1a} \MM^{ab} \epsilon_b^T \\
&\rightarrow& \OO(1/z) \nn \ee where we have used the result from
\cite{ArkaniHamed:2008yf} to throw out the contribution from the
pure gauge theory, $\MM_{\rm gluon}$. By convention gluon 1 can
always be chosen to have negative helicity, so $\MM^{-,+}$,
$\MM^{-,-}$, and $\MM^{-,T}$ characterize any amplitude with two
gluons. Since this amplitude vanishes at large $z$, the recursion
relations hold.

\begin{figure}[t] \be
\parbox{55mm} {
\begin{fmfgraph*}(50,30)
\fmfbottom{p1,p2} \fmftop{k1,k2,k3,k4} \fmf{photon}{p1,q1}
\fmf{photon}{q4,p2}
\fmf{fermion,label=$\sslash{q}$}{q1,q2}\fmf{fermion,label=$\sslash{q}$}{q2,q3}
\fmf{fermion,label=$\sslash{q}$}{q3,q4} \fmf{fermion}{k1,q1}
\fmf{fermion}{k4,q4} \fmffreeze \fmf{photon}{k2,q2}
\fmf{photon}{k3,q3}
\end{fmfgraph*}} \nn
\ee \caption{An example of a diagram with only hard fermion
propagators.  Naively, the leading $z$ contribution goes as
$\OO(1)$ and comes from taking a $q$ from every propagator
numerator. However, after dotting into the external polarization
for particle 1, we find that every such diagram vanishes.}
\end{figure}
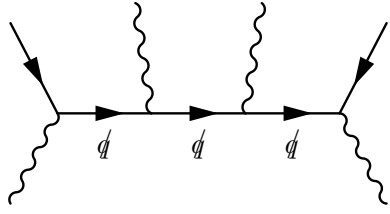

\subsection{Mixed Gluon Amplitudes}

So far we have shown that recursion relations exist for amplitudes
with at least two external gluons. As it turns out, this statement
actually holds more generally, in particular for amplitudes with
only a single external gluon. To see this, consider an amplitude
where particle 1 is a gluon and particle 2 is a scalar or fermion.
As before, we can fix all of the interactions to be $\OO(1)$ using
background field light-cone gauge and the appropriate $R_\xi$
gauge. However, since particle 2 is no longer a gluon, the
explicit diagrams that we must check are now different.

 First, let us consider all diagrams with only hard fermion propagators.
As before, the leading
 order in $z$ contribution comes from taking a $\sslash{q}$
 from every fermion propagator numerator.  No matter whether particle 2 is a scalar or a fermion,
 these diagrams take the form
\be \MM^{a \ldots} &\sim& \gamma^{a} \sslash{q}\ldots \ee
where the hard gluon polarization is dotted into $a$.  This is
always the case because particle 1 has to connect to two fermion
lines, and since it is a gluon this coupling has to be a gauge
interaction. Since we can choose the gluon to have negative
helicity, $\epsilon^-_a = q_a$, this leading order contribution
vanishes. Thus, the contribution from diagrams with only hard
fermion propagators starts at $\OO(1/z)$.

 This leaves diagrams with no hard propagators
(see figures 2d and 2e).  Both mixed diagrams arise from the
$\LL_{(1)}=\tr a_\mu J^\mu_{(1)}$, which naively contains terms
that mix the hard gluon with the derivative of a hard scalar or a
hard fermion. However, our choice of $R_\xi$ gauge eliminates
these  terms and so the diagrams go only as $\OO(1)$. If we then
dot the diagrams into the gluon polarization, $\epsilon^-_a =
-p_a/z$, we find that the diagram vanishes if particle 2 is a
scalar (since it has no polarization to introduce any additional
factors of $z$), but if particle 2 is a fermion, the diagram can
still go as $\OO(1)$.  An explicit check of this fermion diagram
is straightforward, since it is simply a single vertex Feynman
diagram.  After dotting into polarizations, we find that the
leading contribution goes as $\OO(1/z)$, for any fermion
polarization. Thus, we have shown that $\MM(z\rightarrow
\infty)=0$ even when particle 2 is scalar or fermion. This
completes our proof that any amplitude with at least one gluon
obeys a recursion relation.

\section{Spin $\leq 2$ Amplitudes}

In this section we consider a generic theory of gravity coupled to
spin $0, \frac{1}{2},1,\frac{3}{2}$ fields.  Our procedure will
mirror that of the spin $\leq 1$ case.  We find that
$\MM(z\rightarrow \infty) =0$ for any amplitude with at least one
graviton, so this amplitude obeys a recursion relation. Our result
holds for ($\NN=8$) supergravity.

\subsection{Background Field Lagrangian}

Consider (super)gravity coupled to a two derivative theory of
matter (spin $\leq 1$) fields:
\be \LL &=& \sqrt{-g} R + \LL_{\rm
matt}(\Phi,\Psi_\alpha,A_\mu,\Lambda_{\mu\alpha},g_{\mu\nu}) \ee
We expand the action in terms of hard fluctuations around a soft
background
\be g_{\mu\nu} &\rightarrow& g_{\mu\nu} + h_{\mu\nu} \\
\Lambda_{\mu\alpha} &\rightarrow& \Lambda_{\mu\alpha} +
\lambda_{\mu\alpha} \nn\\
A_\mu &\rightarrow& A_\mu + a_\mu \nn\\
\Psi_\alpha &\rightarrow& \Psi_\alpha + \psi_\alpha \nn\\
\Phi &\rightarrow& \Phi + \phi \nn \ee In powers of the graviton
fluctuation, the action becomes
\be \LL &=& \LL_{(0)} + \LL_{(1)}+ \LL_{(2)}+\ldots  \\
\LL_{(2)} &=& \sqrt{-g} \left( \frac{1}{4} \grad_\rho h_{\mu\nu}
\grad^\rho h^{\mu\nu} - \frac{1}{4} \grad_\mu h \grad^\mu h
+\frac{1}{2} \grad_\mu h \grad_\nu h^{\mu\nu} -\frac{1}{2}
\grad_\rho h_{\mu\nu} \grad^\mu h^{\nu\rho} \right. \nn
\\
&& \qquad \;\;\; \left. +\frac{1}{2} h_{\mu\nu} h_{\rho\sigma}
X^{\mu\nu\rho\sigma} + \grad_\lambda h_{\mu\nu} h_{\rho\sigma} Y^{\lambda \mu\nu\rho\sigma} \right) \nn \\
\LL_{(1)} &=& \sqrt{-g} \left(\frac{1}{2} h_{\mu\nu}
T^{\mu\nu}_{(1)}\right) \nn\ee
Here $X^{\mu\nu\rho\sigma}$ and $Y^{\lambda \mu\nu\rho\sigma}$ are
functions of the graviton and matter backgrounds and
$T^{\mu\nu}_{(1)}$ is the stress-energy tensor expanded to linear
order in the matter fluctuations.  For now we will ignore the
precise form of $X^{\mu\nu\rho\sigma}$ and
$Y^{\lambda\mu\nu\rho\sigma}$, but return to them in a later
section.  Also, we have assumed that the background fields obey
their equations of motion, so the fluctuations do not have
tadpoles.

\subsection{Eliminating $\OO(z^2)$ and $\OO(z)$ Vertices}

As in the gauge theory case, our first concern will be
interactions that involve derivatives acting on the hard fields.
In particular, we have to worry about mixing terms between the
graviton and the derivatives of matter fields, which arise from
$\LL_{(1)}$. However, these dangerous contributions can always be
eliminated by an appropriate $R_\xi$ gauge. For example, consider
a free scalar coupled to gravity:
\be \LL_{(1)} = \sqrt{-g}\left(\frac{1}{2} h_{\mu\nu}
T^{\mu\nu}_{(1)} \right)&=& \sqrt{-g} h_{\mu\nu}\left(g^{\mu\rho}
g^{\nu\sigma} - \frac{1}{2}g^{\mu\nu}
g^{\rho\sigma}\right)\grad_\rho \phi \grad_\sigma \Phi \\
&\overset{IbP}{=}& - \sqrt{-g}\left(\grad^\mu h_{\mu\nu} -
\frac{1}{2} \grad_\nu h\right) \phi \grad^\nu \Phi + \ldots \nn\ee
Since this term includes a derivative acting on a hard field, it
will naively introduce $z$'s into amplitudes. However, if we
choose a (deDonder) $R_\xi$ gauge term \cite{'tHooft:1974bx}
\be \LL_\xi &=&  \frac{\sqrt{-g}}{2\xi}\left(\grad^\mu h_{\mu\nu}
- \frac{1}{2} \grad_\nu h + \xi \phi \grad_\nu \Phi \right)^2 \ee
which for $\xi=1$ eliminates these dangerous terms. Since this
gauge choice is simply unitary gauge for the graviton, we know
that it will still work if there are also fermions and gluons in
the theory. After sending $\LL \rightarrow \LL + \LL_\xi$, we find
that the quadratic graviton action becomes
\be \LL_{(2)} &=& \sqrt{-g} \left( \frac{1}{4} \grad_\rho
h_{\mu\nu} \grad^\rho h^{\mu\nu} -\frac{1}{8}\grad_\mu h \grad^\mu
h + \frac{1}{2} h_{\mu\nu} h_{\rho\sigma} X^{\mu\nu\rho\sigma} +
\grad_\lambda h_{\mu\nu} h_{\rho\sigma}
Y^{\lambda\mu\nu\rho\sigma} \right) \ee
As written, the above Lagrangian does not have a manifest spin
Lorentz symmetry.  However, this can be rectified using a trick
from \cite{Bern:1999ji}, whereby a dilaton $\chi$ is introduced
simply to remove the $\grad_\mu h \grad^\mu h$ kinetic term. Then,
we perform a field redefinition
\be \label{eq:redef}h_{\mu\nu} \rightarrow h_{\mu\nu} + g_{\mu\nu}
\sqrt{\frac{2}{D-2}}\chi, &&\qquad \chi \rightarrow \frac{1}{2}
g^{\mu\nu} h_{\mu\nu} + \sqrt{\frac{D-2}{2}}\chi \ee
Because of how the dilaton couples to matter, this field
redefinition effectively eliminates any coupling between matter
and the trace of the graviton, $h$.  Moreover, since dilaton
number is conserved, the dilaton completely decouples from any
tree-level Feynman diagram that does not have external dilaton
legs. We will be concerned only with such diagrams. After the
field redefinition the quadratic graviton action takes the form
\be \LL_{(2)} &=& \sqrt{-g} \left( \frac{1}{4} \grad_\rho
h_{\mu\nu} \grad^\rho h^{\mu\nu} + \frac{1}{2} h_{\mu\nu}
h_{\rho\sigma} X^{\mu\nu\rho\sigma} + \grad_\lambda h_{\mu\nu}
h_{\rho\sigma} Y^{\lambda\mu\nu\rho\sigma} \right) \ee
The combination of the $R_\xi$ gauge and the field redefinition
now makes the spin Lorentz symmetry manifest. To see this, let us
rewrite the action in terms of a left and right vierbein, $e$ and
$\bar{e}$ and a left and right connection, $\omega$ and
$\bar\omega$:
\be g_{\mu\nu} &=& e_\mu^a e_\nu^{b} \eta_{ab}
=\bar{e}_\mu^{\bar{a}} \bar{e}_\nu^{\bar{b}}
\eta_{\bar{a}\bar{b}} \\
h_{\mu\nu} &=& e_\mu^a \bar{e}_\nu^{\bar{a}} h_{a \bar{a}} \nn\\
\grad_\rho h_{\mu\nu} &=& e_\mu^a \bar{e}_\nu^{\bar{a}}
\grad_\rho h_{a \bar{a}} \nn\\
\grad_\rho h_{a \bar{a}} &=& \partial_\rho h_{a \bar{a}}
+\omega^{\;\;b}_{\rho \;\; a} h_{b \bar{a}} +
\bar\omega^{\;\;\bar{b}}_{\rho \;\; \bar{a}} h_{a \bar{b}}  \nn\ee
Of course there is still only a single diffeomorphism redundancy,
and the left/right distinction is introduced only to emphasize
that there are two copies of the spin Lorentz symmetry acting on
the barred and unbarred indices of the graviton. We can see this
because the action now takes the form \be \LL_{(2)}&=& \sqrt{-g}
\left( \frac{1}{4} g^{\lambda\kappa}\eta^{ab}
\eta^{\bar{a}\bar{b}} \grad_\lambda h_{a\bar{a}} \grad_\kappa h_{b
\bar{b}} + \frac{1}{2} h_{a\bar{a}} h_{b\bar{b}}
X^{a\bar{a}b\bar{b}} + \grad_\lambda h_{a\bar{a}} h_{b\bar{b}}
Y^{\lambda a\bar{a}b\bar{b}} \right)\ee
At high energies, the leading contribution is proportional to
$\eta^{ab} \eta^{\bar{a}\bar{b}}$, which enjoys a double spin
Lorentz symmetry.

Next, let us choose light-cone gauge for the background graviton
field, i$.$e$.$ for the metric $g_{\mu\nu}$. This will eliminate
large $z$ contributions coming from derivatives acting on the
background.  As in \cite{ArkaniHamed:2008yf}, we can choose a
gauge in which $q_\mu$ is always in the direction of the negative
helicity polarization, so
%
%
\be g^{--} &=& g^{-i} = \omega^-_{\;\;ab} = \bar\omega^-_{\;\;\bar{a}\bar{b}} =0  \\
g^{-+} &=& 1 \nn  \ee where $i$ labels all directions orthogonal
to the $\pm$ polarizations, that is, the $p_{1,2}$ and
$\epsilon^T$ directions.  This gauge choice ensures that $q$ is
the same as the negative helicity polarization vector in the local
Lorentz frame, i$.$e$.$ $q^2= q \cdot p_{1,2}  = q \cdot
\epsilon^T = 0$, and $q \cdot q^*=1$, where contractions are with
respect $\eta^{ab} = \eta^{\bar{a}\bar{b}}$. This background
light-cone gauge eliminates all large $z$ contributions from
derivative interactions except in the unique diagrams, which were
shown to vanish in \cite{ArkaniHamed:2008yf}.

Just as in the gauge case, any diagram with at least one boson
propagator will get an additional factor of $1/z$, and thus will
vanish at large $z$.  All that is left to check is diagrams with
no hard propagators, and diagrams with only hard fermion
propagators.

\subsection{What are $X^{a\bar{a}b\bar{b}}$  and
$Y^{\lambda a\bar{a}b\bar{b}}$?}

Before checking the remaining diagrams, it will be useful to first
determine $X^{a\bar{a}b\bar{b}}$ and $Y^{\lambda
a\bar{a}b\bar{b}}$.  To begin, one might ask why higher derivative
terms like $\grad_\lambda \grad_\kappa h_{a\bar{a}} h_{b\bar{b}}$
are absent from the graviton action. To see this, observe that the
graviton enters $\LL_{\rm matt}$ only through the metric
(contributing to $X^{a\bar{a}b\bar{b}}$), while derivatives of the
graviton enter only through covariant derivatives (contributing to
$Y^{\lambda a\bar{a}b\bar{b}}$). Since we assume a two derivative
action, the only covariant derivatives come from the kinetic
terms. For spin $0$ and $1$, $\partial_\mu\phi = \grad_\mu \phi$
and $\grad_{[\mu} a_{\nu]} =\partial_{[\mu} a_{\nu]}$, so the
covariant derivatives are simply partial derivatives and thus
bosons do not contribute to $Y^{\lambda a\bar{a}b\bar{b}}$! For
spin $\frac{1}{2}$ and $\frac{3}{2}$, there is a single covariant
derivative which introduces a derivative acting on the graviton:
thus, $Y^{\lambda a\bar{a}b\bar{b}}$ gets contributions from the
fermion kinetic terms and $X^{a\bar{a}b\bar{b}}$ gets
contributions from everything else.

It is possible to deduce the form of $X^{a\bar{a}b\bar{b}}$ and
$Y^{\lambda a\bar{a}b\bar{b}}$ simply by combining spurions made
up from the background fields.  Let us begin with
$X^{a\bar{a}b\bar{b}}$, which has a manifest $a\bar{a}
\leftrightarrow b\bar{b}$ symmetry because it couples to $h_{a
\bar{a}}h_{b \bar{b}}$.  Since $X^{a\bar{a}b\bar{b}}$ has indices,
any background field spurion from which it is constructed must
also have indices.  These spurions come from the kinetic terms:
\be \partial^a \Phi\partial^b \Phi ,\quad \bar{\Psi} \gamma^a
\partial^b \Psi ,\quad F^{ab} F^{cd},\quad
\bar{\Lambda}^a \gamma^{bcd}
\partial^e \Lambda^f,\quad R^{abcd} \ee
where we have not specified whether indices are left or right.
Next, we simply combine these spurions with the metric in order to
obtain a four index tensor. For example, since the spin $0$ and
$\frac{1}{2}$ spurions only have two indices, we simply multiply
them by $\eta^{ab}$ or $\eta^{\bar{a}\bar{b}}$.  Thus, the scalar
contributes the four index tensors $\eta^{ab}
\partial^{\bar{a}}\Phi
\partial^{\bar{b}}\Phi$ and $ \eta^{\bar{a}\bar{b}}\partial^{a}\Phi
\partial^{b}\Phi$.  We might in principle have
multiplied by $\eta^{a \bar{a}}$, but the field redefinition in
equation \ref{eq:redef} eliminates any couplings of matter to the
trace of the graviton, so this contribution vanishes. On the other
hand, the spin $1$ and $2$ spurions have exactly four indices, so
they contribute $F^{ab} F^{\bar{a}\bar{b}}$ and
$R^{ab\bar{a}\bar{b}}$. Lastly, the spin $\frac{3}{2}$ spurion has
six indices and so must be contracted with a metric to yield a
four index tensor. Any single contraction will leave at least two
anti-symmetric indices. Since there is a $a\bar{a} \leftrightarrow
b \bar{b}$ symmetry, this contribution is anti-symmetric in both
$ab$ and $\bar{a}\bar{b}$.  Consequently $X^{a\bar{a}b\bar{b}}$ is
of the form:
\be \label{eq:X} X^{a\bar{a}b\bar{b}} &=&
A^{(ab)}\eta^{\bar{a}\bar{b}} +
\bar{A}^{(\bar{a}\bar{b})}\eta^{ab} + B^{[ab] [\bar{a}\bar{b}]}
\ee
where $B^{[ab][\bar{a}\bar{b}]}$ has the symmetries of the Riemann
tensor.

In contrast, $Y^{\lambda a\bar{a}b\bar{b}}$ only receives
contributions from covariant derivatives in the fermion kinetic
terms.  This is because $\grad_\mu = \partial_\mu +\omega_{\mu ab}
\gamma^{ab}/8$, and when linearized $\omega_{\mu ab}$ contains the
derivative of the graviton.  Moreover, without loss of generality
we can take $Y^{\lambda a\bar{a}b\bar{b}}$ to be odd under
$a\bar{a} \leftrightarrow b\bar{b}$ because the even part can
always be integrated by parts and absorbed into $X^{
a\bar{a}b\bar{b}}$.
%
%
%
%
Before linearizing, $\omega_{\mu ab}$ enters the action as:
\be \LL &\supset & i \omega_{\mu ab} \left( \bar{\Psi} \gamma^\mu
\gamma^{ab} \Psi + \frac{1}{12} \bar{\Lambda}_\rho \gamma^{\rho
\sigma \mu} \gamma^{ab} \Lambda_\sigma \right) \ee  $Y^{\lambda
abcd}$ has to be constructed from a spurion with the same tensor
structure as the quantity in parentheses.  Let us call this
spurion $C^{\mu ab}$, where the $ab$ indices are anti-symmetrized.
Since $Y^{\lambda a\bar{a}b\bar{b}}$ is odd under
$a\bar{a}\leftrightarrow b \bar{b}$, anti-symmetry in $ab$ implies
symmetry in $\bar{a}\bar{b}$ and vice versa.  Thus, in order to
construct a four index tensor with the right symmetry properties,
we multiply the spurion by $\eta^{ab}$ or $\eta^{\bar{a}\bar{b}}$.
Thus $Y^{\lambda a\bar{a}b\bar{b}}$ is
\be \label{eq:Y} Y^{\lambda a\bar{a}b\bar{b}} &=&
C^{\lambda[ab]}\eta^{\bar{a}\bar{b}} +
\bar{C}^{\lambda[\bar{a}\bar{b}]}\eta^{ab} \ee
To see some explicit formulae for $X^{a\bar{a}b\bar{b}}$ and
$Y^{\lambda a\bar{a}b\bar{b}}$, see appendix A. That said, only
the generic structure shown above will be necessary for showing
vanishing large $z$ behavior.

\subsection{Checking Explicit Diagrams}

In this section we show that all diagrams with no hard propagators
and all diagrams with only hard fermion propagators vanish at
large $z$. For now, consider the case where particles 1 and 2 are
both gravitons.

Diagrams with no hard propagators contain only one interaction
vertex involving hard momenta, so their contribution to the
amplitude can be read directly off the action:
\be \delta \MM_1^{a\bar{a} b\bar{b}} &=& X^{a\bar{a}b\bar{b}} +
(p_1 + zq)_\lambda Y^{\lambda a\bar{a}b \bar{b}} \\
&=& A^{ab}\eta^{\bar{a}\bar{b}} +
\bar{A}^{\bar{a}\bar{b}}\eta^{ab} + B^{[ab] [\bar{a}\bar{b}]}+ z
(C^{[ab]}\eta^{\bar{a}\bar{b}} +
\bar{C}^{[\bar{a}\bar{b}]}\eta^{ab} ) \nn\ee
where $A^{ab} = A^{(ab)} + p_{1\lambda}C^{\lambda ab}$ and
$q_\lambda C^{\lambda ab} = C^{ab}$ and the same for the barred
variables.  It is interesting to note that $\delta
\MM_1^{a\bar{a}b\bar{b}}$ has precisely the same tensor structure
as the pure gravity amplitude computed in
\cite{ArkaniHamed:2008yf}.

Next, let us consider diagrams with only hard fermion propagators.
For the most part these diagrams are the same as in the spin $\leq
1$ case. The only subtlety is that the propagator for a spin
$\frac{3}{2}$ fermion is different from that of a spin
$\frac{1}{2}$ fermion. The kinetic term for the gravitino
fluctuation is:
\be \LL_{\frac{3}{2}} &=& \frac{i}{12} \bar{\lambda}_\rho
\gamma^{\rho\sigma\mu} \grad_\mu \lambda_\sigma \ee
However, by choosing a gauge $\gamma^\mu \lambda_\mu =0$ and
anti-commuting gamma matrices, we obtain
\be \LL_{\frac{3}{2}} &=& -\frac{i}{2} \bar{\lambda}_\mu
\sslash{\grad} \lambda^\mu \ee
which is simply four copies of a spin $\frac{1}{2}$ fermion.
Notice the manifest spin Lorentz symmetry acting on gravitino
index!

Like before, the leading in $z$ contribution comes from taking a
factor of $\sslash{q}$ from every fermion propagator numerator.
Applying the same arguments as in the gauge theory case, the
contribution from diagrams with only hard fermion propagators
becomes
\be \delta \MM^{a\bar{a}b\bar{b}}_2 =
D^{[ab]\bar{a}\bar{b}}+\bar{D}^{[\bar{a}\bar{b}]ab} &+&
q^{(a}E^{b)\bar{a}\bar{b}}-\eta^{ab} q_c E^{c \bar{a}\bar{b}}
\\
&+& q^{(\bar{a}}\bar{E}^{\bar{b})ab}-\eta^{\bar{a}\bar{b}}
q_{\bar{c}} \bar{E}^{\bar{c} ab} \nn\ee
which is very similar to the corresponding expression in the gauge
theory (see equation \ref{eq:fermion}) except that there are two
additional indices.

Summing contributions from all diagrams with no hard propagators
and all diagrams with only hard fermion propagators, we obtain the
amplitude
\be \label{eq:gravity} \MM^{a\bar{a}b\bar{b}} = \MM_{\rm
grav}^{a\bar{a}b\bar{b}} &+& \delta \MM_1^{a\bar{a}b\bar{b}} +
\delta \MM_2^{a\bar{a}b\bar{b}}
\\ = \MM_{\rm grav}^{a\bar{a}b\bar{b}} &+& A^{ab}\eta^{\bar{a}\bar{b}} +
\bar{A}^{\bar{a}\bar{b}}\eta^{ab} + B^{[ab] [\bar{a}\bar{b}]}+ z
(C^{[ab]}\eta^{\bar{a}\bar{b}} +
\bar{C}^{[\bar{a}\bar{b}]}\eta^{ab} ) \nn\\ &+&
D^{[ab]\bar{a}\bar{b}}+\bar{D}^{[\bar{a}\bar{b}]ab} +
q^{(a}E^{b)\bar{a}\bar{b}}-\eta^{ab} q_c E^{c \bar{a}\bar{b}} +
q^{(\bar{a}}\bar{E}^{\bar{b})ab}-\eta^{\bar{a}\bar{b}} q_{\bar{c}}
\bar{E}^{\bar{c} ab} + \OO(1/z) \nn\ee
where $\MM_{\rm grav}^{a\bar{a}b\bar{b}}$ is the contribution from
pure gravity considered in \cite{ArkaniHamed:2008yf}. Without loss
of generality, we take graviton 1 to have negative helicity and
graviton 2 to be arbitrary.  The graviton polarizations are
symmetric, traceless products of gauge polarizations, so they take
the form
\be \epsilon_{1a\bar{a}}^{--} &=& q_a q_{\bar{a}} \overset{\rm gauge}{=} p_{1a}p_{1\bar{a}}/z^2 \\
\epsilon_{2a\bar{a}}^{\pm\pm} &=& \left\{ \begin{array}{cc} q_a
q_{\bar{a}}, &
\textrm{$(++)$} \\
(q_{a}^*+z p_{1a})(q_{\bar{a}}^*+z p_{1\bar{a}}), &
\textrm{$(--)$}
  \end{array}
  \right.\nn
\ee
Using equation \ref{eq:gravity}, we find that the $(--,++)$,
$(--,--)$, and $(--,T)$ amplitudes go as:
\be \MM^{--,++} &=& q_a q_{\bar{a}} \MM^{a\bar{a}b\bar{b}}q_b
q_{\bar{b}}
\\ &\rightarrow& \OO(1/z) \nn \\
\MM^{--,--}&=& \frac{1}{z^2}p_{1a}p_{1\bar{a}}
\MM^{a\bar{a}b\bar{b}}(q_b^*+z p_{1b})(q_{\bar{b}}^*+z
p_{1\bar{b}}) \\
&\rightarrow& \OO(1/z) \nn\\
\MM^{--,T} &=& \frac{1}{z^2}p_{1a}p_{1\bar{a}}
\MM^{a\bar{a}b\bar{b}}\epsilon_b^T \epsilon_{\bar{b}}^T \\ &
\rightarrow& \OO(1/z^2) \nn\ee
Thus, we have shown all amplitudes with at least two gravitons
vanish at large $z$.


\subsection{Mixed Graviton Amplitudes}

In theories of (super)gravity coupled to scalars and fermions, any
tree amplitude with at least two gravitons obeys a recursion
relation.  However, in analogy with the gauge case, gravitational
theories also admit recursion relations for amplitudes with only
one graviton. Our argument parallels that of the spin $\leq 1$
theory.  Again, by fixing light-cone gauge for the background and
an $R_\xi$ gauge for the fluctuation we can remove all $\OO(z^2)$
and $\OO(z)$ interactions.  Then we just have to check explicit
diagrams in which particle 1 is a graviton and particle 2 is not.

First, let us consider diagrams with only hard fermion
propagators. No matter the identity of particle 2, these diagrams
take the form
\be \MM^{a\ldots} &\sim& \gamma^{a} \sslash{q} \ldots \ee
where $a$ is a graviton index.  This has to be true because
particle 1 connects to two fermion lines, and gravitons can only
couple to fermions in a very specific way. Finally, dotting
$\MM^{a\ldots}$ into a negative helicity graviton polarization
$\epsilon^{--}_{1a\bar{a}}= q_a q_{\bar{a}}$, we see that this
leading in $z$ contribution vanishes.

Diagrams with no hard propagators come from the term $ h_{\mu\nu}
T^{\mu\nu}_{(1)}$, which naively includes interactions involving
derivatives of hard matter fields. However, the deDonder $R_\xi$
gauge removes such terms. Consequently, the interaction vertex
goes at most as $\OO(1)$. Next, dotting the mixed diagram into the
 graviton polarization
$\epsilon^{--}_{1a\bar{a}} = p_a p_{\bar{a}} /z^2$, we realize
that if particle 2 is a scalar, fermion, or gluon, then its
polarization goes at most as $z$ and so this contribution to the
amplitude vanishes at large $z$. Finally, by explicitly checking
the diagram where particle 2 is a gravitino, we find that all of
the mixed diagrams vanish at large $z$. This completes our proof
that recursion relations hold for any amplitude with at least one
graviton.

\section{Conclusion}

Recursion relations are a generic feature of tree amplitudes that
vanish at large complexified momentum. In this paper we show that
this criterion holds for a broad class of amplitudes in two
derivative gauge and (super)gravity theories. In particular, for a
theory of spin $\leq 1$, any amplitude with at least one gluon can
be recursed; for a theory of spin $\leq 2$ this is true of any
amplitude with at least one graviton.  Said another way, recursion
relations hold as long there is at least one external leg with the
highest spin possible. This is sensible because a higher spin
particle enjoys a greater gauge redundancy that is essential for
obtaining nice large $z$ behavior in an amplitude.  In particular,
only by choosing light-cone gauge for the backgrounds and an
additional $R_\xi$ gauge for the fluctuations were we able to
eliminate large $z$ contributions. Moreover, the $R_\xi$ gauge was
especially critical for proving vanishing large $z$ behavior in
amplitudes with only one gluon or only one graviton.

Finally, we remark on the interesting fact that the spin $\leq 2$
amplitude (equation \ref{eq:gravity}) has precisely the same
structure as the square of the spin $\leq 1$ amplitude (equation
\ref{eq:gauge}). This is a non-trivial consistency check against
the famous KLT relation \cite{Kawai:1985xq} that equates closed
string tree amplitudes with sums over products of open string tree
amplitudes.  At low energies this statement persists as a relation
between amplitudes in Yang-Mills and gravity, and in fact for the
case of MHV, new formulas for the the KLT relations have been
derived directly from the BCFW recursion relations
\cite{Elvang:2007sg}. Since the KLT relations also relate
amplitudes in $\NN =4$ SYM and $\NN=8$ supergravity (a subset of
the theories under consideration in this paper), we should expect
such a relation between our expressions for spin $\leq 1$ and spin
$\leq 2$ amplitudes.

\acknowledgments{It is a pleasure to thank N. Arkani-Hamed, H.
Elvang, D. Freedman, and J. Kaplan for very helpful physics
discussions and comments on the manuscript. CC is supported in
part by DOE grant DE-FG02-91ER40654.}

\appendix

\section{Exact Expressions for $X^{\mu\nu\rho\sigma}$
and $Y^{\lambda \mu\nu\rho\sigma}$} \label{appendix:XY}

In this section we write down some exact expressions for
$X^{\mu\nu\rho\sigma}$ and $Y^{\lambda \mu\nu\rho\sigma}$. First,
let us parse these tensors according to the spin of the field
contributing: $s=0,\frac{1}{2},2$. An explicit calculation shows
\cite{BjerrumBohr:2004mz}:
\be X^{\mu\nu\rho\sigma}_{s=0} &=& \eta^{\mu\rho} \partial^\nu
\Phi
\partial^\sigma \Phi - \frac{1}{2} \eta^{\mu\nu} \partial^\rho\Phi
\partial^\sigma \Phi - \frac{1}{4} \eta^{\mu\rho} \eta^{\nu\sigma}
\partial_\lambda\Phi \partial^\lambda\Phi + \frac{1}{8} \eta^{\mu\nu} \eta^{\rho\sigma}
\partial_\lambda\Phi \partial^\lambda\Phi \\
X^{\mu\nu\rho\sigma}_{s=\frac{1}{2}} &=& \frac{3}{8}
\eta^{\mu\rho}\bar{\Psi} i \gamma^\nu \widehat{\grad}^\sigma \Psi
-\frac{1}{4} \eta^{\mu\nu} \bar{\Psi} i \gamma^\rho
\widehat{\grad}^\sigma \Psi
-\frac{1}{4}\eta^{\mu\rho}\eta^{\nu\sigma} \bar{\Psi} i
\widehat{\sslash{\grad}} \Psi
+\frac{1}{8}\eta^{\mu\nu}\eta^{\rho\sigma} \bar{\Psi} i
\widehat{\sslash{\grad}} \Psi \nn\\
 X^{\mu\nu\rho\sigma}_{s=2} & =& -R^{\mu\rho \nu\sigma}+2
\eta^{\mu\rho}R^{\nu\sigma} + \eta^{\mu\nu} R^{\rho\sigma} +
\frac{1}{2} \eta^{\mu\rho} \eta^{\nu\sigma} R - \frac{1}{4}
\eta^{\mu\nu}\eta^{\rho\sigma} R \nn \\
Y^{\lambda\mu\nu\rho\sigma}_{s=0} &=& 0 \nn \\
Y^{\lambda\mu\nu\rho\sigma}_{s=\frac{1}{2}} &=& \frac{1}{48}
g^{\nu\sigma} \bar{\Psi} \gamma^{\mu\rho\lambda} \Psi \nn \\
Y^{\lambda\mu\nu\rho\sigma}_{s=2} &=&0 \nn  \ee where
$\bar{\Psi}\widehat{\grad}_\mu \Psi=\bar{\Psi}\grad_\mu
\Psi-\grad_\mu\bar{\Psi} \Psi$ and we have only included
contributions from the kinetic terms.  Aside from terms that
couple to the trace of the graviton (which are removed by the
dilaton field redefinition) these expressions match the general
structure deduced in equations \ref{eq:X} and \ref{eq:Y}.


\end{fmffile}

\end{document}